\documentclass[floats,floatfix,showpacs,preprint,prd,superscriptaddress,onecolumn,nofootinbib]{revtex4}
\usepackage{graphicx, epsfig, amssymb} 
\usepackage{amsmath, amsfonts}
\usepackage{bm} 
\usepackage{color}
\usepackage[usenames]{xcolor}

\newcommand{\non}{\nonumber}
\newcommand{\be}{\begin{equation}}
\newcommand{\ee}{\end{equation}}
\newcommand{\bea}{\begin{eqnarray}}
\newcommand{\eea}{\end{eqnarray}}

\begin{document}

\title{Static and rotating solutions for Vector-Galileon theories}

\author{Adolfo Cisterna}
\email{adolfo.cisterna@ucentral.cl}
\affiliation{Universidad Central de Chile, Santiago, Chile}
\affiliation{Instituto de Ciencias F\'isicas y Matem\'aticas, Universidad Austral de Chile,\\ Casilla 567,
Valdivia, Chile}

\author{Mokhtar Hassaine}
\email{hassaine@inst-mat.utalca.cl}
\affiliation{Instituto de Matem\'{a}tica y F\'{\i}sica, Universidad de Talca, Casilla
747, Talca, Chile}

\author{Julio Oliva}
\email{juoliva@udec.cl}
\affiliation{Instituto de Ciencias F\'isicas y Matem\'aticas, Universidad Austral de Chile,\\ Casilla 567,
Valdivia, Chile}
\affiliation{Departamento de F\'isica, Universidad de Concepci\'on, Casilla, 160-C,\\ Concepci\'on, Chile}

\author{Massimiliano Rinaldi}
\email{massimiliano.rinaldi@unitn.it}
\affiliation{Dipartimento di Fisica, Universit\`a di Trento,Via Sommarive 14,\\ 38123 Povo (TN), Italy}
\affiliation{TIFPA - INFN,\\  Via Sommarive 14, 38123 Povo (TN), Italy}

\begin{abstract}

We consider a particular truncation of the generalized Proca field theory in four dimensions for which we construct a static and axisymmetric rotating black hole
``stealth solutions'', namely solutions with (Anti) de Sitter or Kerr metric but non-trivial vector field. The static configuration is promoted to a backreacting black hole with asymptotic (Anti) de Sitter behavior by turning on a nonlinear electrodynamic source given as a fixed power of the Maxwell invariant. Finally we extend our solutions to arbitrary dimensions.

\end{abstract}


\maketitle

\date{\today}

\section{Introduction}

So far, the most successful theory of gravity, not only from a theoretical point of view
but also from its experimental validation, is the Einstein's theory of General Relativity (GR).
At Solar System scales, the predictive power of GR \cite{Will:2005va} and the recent direct detection of
gravitational waves \cite{GW} position GR in an unsurpassable place. As a direct consequence,
new horizons on the study of black hole physics will be opened with the opportunity of detecting
black holes via modern observations. Nevertheless, during the last decades, huge efforts
have been made in the construction of alternative theories of gravity \cite{Clifton:2011jh},
mostly motivated by the incompatibilities of GR at quantum scales \cite{Gibbons:1994cg} and by the
dark matter and dark energy phenomena \cite{Sahni:2004ai}. It seems that the theory should be
modified not only at ultraviolet (UV) scales but also at infrared (IR) ones.  Ultraviolet modifications
generally incorporate high order curvature terms as it is the case in String Theory \cite{Polchinski:1998rq},
while IR modifications might imply the inclusion of exotic forms of matter or  new
degrees of freedom. The so far unsuccessful detection of new particles that could  account for  dark matter or
dark energy have motivated many theorists to consider modified gravity with increasingly complex formulations. The most illustrative case is  the so-called Scalar-Tensor Theory (STT), which extends GR with one or more scalar degree of freedom \cite{Brans:1961sx}. A common feature of these theories is that they are all encoded by second-order differential
equations of motion in order to avoid the Ostrogradski instability \cite{Woodard:2015zca}. The most general second-order STT in four dimensions was constructed during the early
seventies by Horndeski \cite{Horndeski:1974wa}, and a sector of it was rediscovered later in a different framework
by the name of Galileon theory \cite{Nicolis:2008in, galileon1, Deffayet:2009mn}. The latter also corresponds to a scalar
field theory coming from the generalization of the decoupling limit of the Dvali-Gabadadze-Porrati (DGP) model \cite{DGP}.
In the last few years, Horndeski/Galileon-like theories have been extensively studied in the context of black hole
physics \cite{Sotiriou:2013qea, rinaldi, Kolyvaris:2011fk, charmousis1, adolfo2, Minam1, Kobayashi:2014eva,
adolfo3, Bravo-Gaete:2013dca, minas, Feng:2015oea, Cisterna:2015uya, Babichev:2015rva, Babichev:2016fbg}. In addition, solutions describing neutron stars and other compact objects have came out, imposing several constraints on the validity of these kinds of models \cite{Cisterna:2015yla, Cisterna:2016vdx, Maselli:2016gxk, Brihaye:2016lin}. These theories attracted a lot of interest also in the cosmology community as they might play a role in explaining inflation, dark energy or dark matter \cite{cosmo}.

 There are other modifications of gravity that are currently under scrutiny as, for example, the Vector-Tensor Theories (VTT). In Ref.~\cite{Deffayet:2013tca}, it  was proved that there is no Galileon
extension for a VTT exhibiting gauge symmetry. In fact, the first result on the general gauge invariant
vector theory coupled to gravity  yielding second-order field equations was obtained by Horndeski himself
\cite{Horndeski:1976gi, Horndeski:1977kz}. The resulting gauge invariant theory, apart from the Maxwell term,
contains an additional contribution proportional to the double dual of the Riemann tensor, and reduces to the standard Maxwell
electrodynamics in the flat limit case. However, relaxing the hypothesis of gauge invariance, more general VTT can
be constructed. The simplest way to do so is to considerer a mass term $m^2A_{\mu}A^{\mu}$ for the vector field, which explicitly
breaks the gauge invariance. The generalization of the Proca action for a massive vector field with derivative self-interaction was studied in several papers, see e.g. \cite{Heisenberg:2014rta, Allys:2015sht, Jimenez:2016isa}. The resulting theory, which describes a massive vector field theory propagating three degrees of freedom, namely the two transverse modes and the longitudinal mode, has been called ``Vector Galileon''. This terminology is essentially due to the fact that the longitudinal mode, which can be related to the scalar
Galileon field through the Stuckelberg mechanism, shows the same self interaction as in the Galileon theory. The corresponding
curved version is obtained using the same procedure as in the scalar case following the standard minimal coupling approach
and adding proper nonminimal couplings between gravity and the vector field in order to maintain the second order nature of the field equations.
As a result, the most general theory describing a Proca field in curved spacetime, yielding second order differential equations of motion and propagating only the three physical degrees of freedom is expressed by the following Lagrangian \cite{Jimenez:2016isa}
\begin{equation}  \label{generalizedProfaField_curved}
\mathcal L^{\rm curved}_{\rm gen. Proca} = \sqrt{-g}\sum^5_{n=2} \mathcal L_n,
\end{equation}
where
\begin{eqnarray}\label{vecGalcurv}
&&\mathcal L_2 =  G_2(A_\mu,F_{\mu\nu},\tilde{F}_{\mu\nu}), \,\, \mathcal L_3  = G_3(Y)\nabla_\mu A^\mu ,\,\,
\mathcal L_4  =  G_{4}(Y)R+G_{4,Y} \left[(\nabla_\mu A^\mu)^2-\nabla_\rho A_\sigma \nabla^\sigma A^\rho\right], \nonumber\\
&&\mathcal L_5 =  G_5(Y)G_{\mu\nu}\nabla^\mu A^\nu-\frac{1}{6}G_{5,Y} \Big[
(\nabla\cdot A)^3 +2\nabla_\rho A_\sigma \nabla^\gamma A^\rho \nabla^\sigma A_\gamma -3(\nabla\cdot A)\nabla_\rho A_\sigma \nabla^\sigma A^\rho \Big]\nonumber \\
&&-\tilde{G}_5(Y) \tilde{F}^{\alpha\mu}\tilde{F}^\beta_{\;\;\mu}\nabla_\alpha A_\beta   \\
&&\mathcal L_6  =  G_6(Y)L^{\mu\nu\alpha\beta}\nabla_\mu A_\nu \nabla_\alpha A_\beta
+\frac{G_{6,Y}}{2} \tilde{F}^{\alpha\beta}\tilde{F}^{\mu\nu}\nabla_\alpha A_\mu \nabla_\beta A_\nu.\nonumber
\end{eqnarray}
Here, $F_{\mu\nu}=\nabla_\mu A_\nu-\nabla_\nu A_\mu$ stands for the field strength tensor,
$\tilde{F}_{\mu\nu}$ its dual and $L^{\mu\nu\alpha\beta}$ is the double dual Riemann tensor defined as
\begin{eqnarray}
L^{\mu\nu\alpha\beta}=\frac14\epsilon^{\mu\nu\rho\sigma}\epsilon^{\alpha\beta\gamma\delta}R_{\rho\sigma\gamma\delta}.
\end{eqnarray}
In the previous expressions, the $G_n$'s represent arbitrary functions of $Y=-\frac{1}{2}A_\mu A^{\mu}$, and we
note that the standard Maxwell term $\frac{1}{4}F_{\mu\nu}F^{\mu\nu}$ may be contained in the function $G_2$.

Recently, several works have appeared studying various aspects of this theory, such as cosmological perturbations \cite{DeFelice:2016yws}, screening mechanisms \cite{DeFelice:2016cri} or higher order extensions \cite{Heisenberg:2016eld}. However, only few works have explored the existence of black hole configurations on the spectrum of these theories. A promising
and interesting sector of \eqref{generalizedProfaField_curved}, which displays black hole solutions with various asymptotic structures, is the one involving the nonminimal coupling of the Proca field with the Einstein tensor, i.e. the term $G_{\mu\nu}A^{\mu}A^{\nu}$, see
\cite{Geng:2015kvs, Fan:2016jnz, Chagoya:2016aar, Minamitsuji:2016ydr}.
\\
Indeed in \cite{Geng:2015kvs}  the author have obtained asymptotically flat and asymptotically Lifshitz black hole solutions. In \cite{Minamitsuji:2016ydr} the author express the relation between solutions on this vector theory and the known solutions for the kinetic nonminimal sector of the scalar case. Moreover the slowly rotating extensions are obtained. On the other hand solutions where a nonminimal coupling between the Ricci scalar and the vector field is considered  are also described in \cite{Fan:2016jnz}.\\

In this paper, we will consider a particular truncation of the general Lagrangian (\ref{generalizedProfaField_curved}) where
the only non-vanishing terms are given by $\mathcal L_3=-\frac{1}{2}A^2\nabla_\mu A^{\mu}$ and by $\mathcal L_2$ which only depends on $F_{\mu\nu}$. Indeed, this truncation
is of interest since as we show below, as a matter of fact, it allows for the construction of black hole solutions. It's also interesting to note that the full theory can be promoted to a gauge invariant theory making use of the Stucckelberg procedure. Indeed, this can be achieved including an additional scalar field through $A_\mu \rightarrow A_\mu + \partial_\mu \pi$.  Properly choosing the arbitrary functions in (\ref{generalizedProfaField_curved}), namely $G_n=Y$, and setting $A_\mu=0$ the full scalar Galileon interactions are recovered. Then it is possible to make a connection between vector and scalar models noticing that our model defined by $\mathcal L_3$ has the DGP model Lagrangian as scalar counterpart. Moreover, due to the fact that $\mathcal L_3$ is linear in the connection its covariantization is trivial and does not need any counterterm when going from Minkowski spacetime to a curved background.

To begin with, we show that, when the term $\mathcal L_2$ is absent in our truncated Lagrangian, the theory admits a particular class of black hole solutions
known as ``stealth configurations''.  These are characterized by the vanishing of  the geometric and of the matter parts of the Einstein equations. In particular, we obtain stealth configurations described by Schwarzschild and Kerr metric while some components of the vector field are non-trivial.

Next, we show that the metric of these stealth configurations can be non-trivially modified by turning on the Lagrangian $\mathcal L_2$. This task is non-trivial essentially because of the non-Coulombian behavior of the potential scalar. Nevertheless, we will take advantage of the nonlinear electrodynamic models that are known to accommodate non-Coulombian fields, and choose the appropriate form for the Lagrangian $\mathcal L_2$, given by a fractional power of the Maxwell invariant.

The paper is organized as follows. In the next section, we present the model and its associated field equations. In Section III, we find the static stealth solution without the Lagrangian $\mathcal L_2$ and the black hole solution when the later is taken into account. In Section IV, we find the stealth configuration corresponding to the Kerr black hole. In the last section, we present some possible extensions of the present work. An appendix is also provided where the four-dimensional stealth configuration on Kerr spacetime is generalized to arbitrary dimensions.

\section{Model and field equations}
We consider the following action defined in four dimensions
\begin{equation}
S=\int{\sqrt{-g}\left[\kappa(R-2\Lambda)-\frac{1}{4}\mathcal{L}_2(F^2)-\frac{\alpha}{2}A^2\nabla_{\nu}A^{\nu}\right]}d^4x,   \label{Lag}
\end{equation}
and corresponding to a subset of the Lagrangian (\ref{generalizedProfaField_curved}-\ref{vecGalcurv}) with  $G_3=-\frac{1}{2}A^2$,
and $\mathcal L_2$ depending  on the Maxwell kinetic term only. The variation of the action with respect to the metric and to the vector field
yield, respectively,
\begin{eqnarray}
E_{\mu\nu}&:=&\kappa(G_{\mu\nu}+\Lambda g_{\mu\nu})-\frac{\alpha}{2}A_{\mu}A_{\nu}\nabla_{\alpha}A^\alpha+\frac{\alpha}{4}A_\mu\nabla_{\nu}(A^2)+\frac{\alpha}{4}A_\nu\nabla_{\mu}(A^2) \nonumber\\
&-&\frac{\alpha}{4}g_{\mu\nu}A^\alpha\nabla_{\alpha}(A^2)-\frac{1}{2}\left(\frac{d\mathcal{L}_2(F^2)}{d F^2}F_{\mu\alpha}F_{\nu}^{^\alpha}-g_{\mu\nu}\mathcal{L}_2(F^2)\right)=0,\label{sys1}
\end{eqnarray}
\begin{equation}
\mathcal{E}^{\nu}:=\nabla_{\mu}\left(F^{\mu\nu}\frac{d\mathcal{L}_2(F^2)}{dF^2}\right)-A^{\nu}\nabla_{\mu}A^{\mu}+\frac{1}{2}\nabla^{\nu}(A^2)=0. \label{sys2}
\end{equation}
Before looking for solutions to the field equations, we first show
that, imposing the following condition
\begin{equation}
\nabla_{\mu}A^{\mu}=0, \label{cond1}
\end{equation}
will imply that, under some reasonable assumptions, the norm of
the vector $A^{\mu}$ must vanish. As shown below, these two
conditions will then be imposed in order to find some solutions.

Indeed, by taking the divergence of the vector field equation, one
finds
\begin{eqnarray*}
\nabla_\nu\nabla_{\mu}\left(F^{\mu\nu}\frac{d\mathcal{L}_2(F^2)}{dF^2}\right)=\nabla_\nu A^{\nu}\nabla_{\mu}A^{\mu}-\frac{1}{2}\nabla_\nu \nabla^{\nu}(A^2).
\end{eqnarray*}
The antisymmetry of the left side of this expression implies that the norm of the vector satisfies the massless
Klein-Gordon equation
\begin{equation}
\Box{A^2}=0.
\label{KGA2}
\end{equation}
Following the same argument used by Bekenstein  to prove its no-hair theorem \cite{Bekenstein:1972ky}, we
now show that, for a spacetime describing a stationary and axisymmetric black hole, equation (\ref{KGA2})
implies $A_{\mu}A^{\mu}=0$, provided that the vector field is regular outside the black hole horizon
and vanishing at infinity. Let us define $\psi=A_{\mu}A^{\mu}$. Since the spacetime is
stationary axisymmetric there must exist a parametrization in which the metric reads
\begin{equation}
ds^{2}=g_{tt}dt^{2}+2g_{t\phi}dtd\phi+g_{\phi\phi}d\phi^{2}+W\left[  d\rho
^{2}+dz^{2}\right]  \ ,
\end{equation}
where the functions $g_{tt}$, $g_{t\phi}$, $g_{\phi\phi}$ and $W$ depend only on $\rho$ and $z$. Assuming that also
the field $\psi$ has the same symmetries\footnote{In the case of a complex Klein-Gordon scalar field with a massive term,
it has been recently shown that relaxing the symmetry condition but still requiring the energy-momentum tensor to realize this symmetry,
hairy black holes exist \cite{Herdeiro:2014goa}. A quite similar argument was previously used to construct non-linear solitons for the Skyrme model \cite{Canfora:2013osa}.} implies that $\psi=\psi\left(  \rho,z\right)$. Then, considering
the equation (\ref{KGA2}) for $\psi$ and integrating over a four-volume $V$ bounded by the horizon, spacelike infinity as well as two spacelike
hypersurfaces at constant $t$, one obtains from the Gauss's theorem that
\begin{eqnarray*}
\int_{t=t_{1}}dS_{\mu}\psi\partial^{\mu}\psi+\int_{t=t_{2}}dS_{\mu}%
\psi\partial^{\mu}\psi+\int_{i^{0}}dS_{\mu}\psi\partial^{\mu}\psi
+\int_{\mathcal{H}^{+}}dS_{\mu}\psi\partial^{\mu}\psi \\\non\\\
-\int_{V}\sqrt{-g}%
d^{4}xW^{-2}\left[  \left(  \psi_{,z}\right)  ^{2}+\left(  \psi_{,\rho
}\right)  ^{2}\right]  =0\;,
\end{eqnarray*}
where we have separated the boundary integral on its different components. The
first two terms corresponding to integrals on spacelike hypersurfaces at
constant $t$ cancel each other because their orientation is opposite and nor
the scalar field neither the metric depend on the time coordinate. The
integral at spacelike infinity ($i^{0}$) vanishes if we require the field
$\psi$ to decay fast enough, and the boundary integral at the
horizon vanishes as well using the Schwarz inequality together with the
fact that the horizon is a null surface. Therefore, the bulk integral must vanish
and, since the integrand is definite positive, the field $\psi$ must be an arbitrary constant. However, the
vanishing condition at infinity implies that
\begin{equation}
A_{\mu}A^{\mu}=0\ .   \label{cond2}
\end{equation}

In the next section, we find static solutions for which both conditions (\ref{cond1})
and (\ref{cond2}) are satisfied.

\section{Two different classes of static solutions}
In this section, we show the construction of static solutions. In order to achieve this task, we consider
the  Ansatz
\begin{equation}
 ds^2=-N(r)^2f(r)dt^2+\frac{dr^2}{f(r)}+r^2\left(d\theta^2+\sin^2\theta\, d\phi^2\right),
\label{metric}
\end{equation}
and
\begin{equation}
A=A_\mu dx^\mu=A_t(r)dt+A_r(r) dr,
\label{proca}
\end{equation}
where the non-trivial component $A_r$, which is related to the vector longitudinal polarization mode, is propagating because of the lack of
gauge symmetry. As mentioned before, the class of solutions we are looking for are those satisfying the conditions (\ref{cond1})
and (\ref{cond2}). It is easy to see that the integration of these two conditions can be done in full generality yielding
\begin{equation}
A_t(r)=\frac{Q}{r^2}, \qquad    A_r(r)=\frac{Q}{r^2\,f(r)\, N(r)},   \label{component0}
\end{equation}
where Q is an arbitrary integration constant. It is important to stress that the apparent singularity of the component $A_r(r)$
is just an artifact that could be removed by using coordinates which are well defined at the horizon.

\subsection{Static black hole stealth configurations}
To determine the metric, we note that when $\mathcal{L}_2=0$, conditions (\ref{cond1})
and (\ref{cond2}) imply that  \eqref{sys2} is automatically satisfied while the matter part of the
stress tensor of the Einstein equations (\ref{sys1}) vanishes. Hence, in order to satisfy both equations, we are left with nothing but Einstein equations with a cosmological constant, $G_{\mu\nu}+\Lambda g_{\mu\nu}=0$,  whose static black hole solutions are given by the
Schwarzschild-(A)dS spacetimes. Hence, the Schwarzschild-(A)dS black hole metric (\ref{metric}) with
$$
N(r)=1,\qquad f(r)=\frac{r^2}{l^2}+1-\frac{2M}{r},
$$
together with  (\ref{component0}), is a solution of the theory defined by the action (\ref{Lag}) with $\mathcal{L}_2=0$.
This model, where both sides of the Einstein equations (the geometric part and the matter stress tensor) vanish identically, is known in the literature as ``stealth'' configuration. Some examples of such solutions have been derived previously in the case of a
scalar field nonminimally coupled to gravity \cite{AyonBeato:2004ig, AyonBeato:2005tu, Ayon-Beato:2015qfa} and also in the context of
Horndeski theories \cite{charmousis1}. It is worth noticing that even though these configurations do not gravitate, in the case of black hole backgrounds
the thermal quantities might depend on the matter profile, which may produce quantum tunneling between the configurations with and without matter field, see e.g. \cite{Maeda:2012tu}.

\subsection{Black hole solutions with nonlinear electrodynamics}

We now  turn on the term $\mathcal{L}_2$ in (\ref{Lag}) in order to find new black hole solutions. Direct integration of the general equations of motion is not possible, thus we opt for the following strategy: we keep imposing the conditions (\ref{cond1}) and (\ref{cond2}) so the solution $A_{\mu}$ is given by (\ref{component0}) with the
static Ansatz (\ref{metric}). In turn, this implies that the field equations (\ref{sys1}-\ref{sys2}) reduce to
\begin{eqnarray}
&&\kappa(G_{\mu\nu}+\Lambda g_{\mu\nu})=\frac{1}{2}\left(\frac{d\mathcal{L}_2(F^2)}{d F^2}F_{\mu\alpha}F_{\nu}^{^\alpha}-g_{\mu\nu}\mathcal{L}_2(F^2)\right),\nonumber\\
\\
&&\nabla_{\mu}\left(F^{\mu\nu}\frac{d\mathcal{L}_2(F^2)}{dF^2}\right)=0.\nonumber
\end{eqnarray}
Now, it remains to find the appropriate Lagrangian $\mathcal{L}_2$ that satisfies the above equations. For sure, because of the non-Coulombian behavior of the scalar potential $A_t(r)=Q/r^2$, we know that the Lagrangian $\mathcal{L}_2$ cannot be given by the standard Maxwell term. A simple way to circumvent
this problem is to choose a form inspired by nonlinear electrodynamics for the Lagrangian $\mathcal{L}_2$. As shown below, the nonlinearity can in fact induce a non-Coulombian scalar potential. It is interesting to note that nonlinear electrodynamics models have been proved to be  excellent laboratories in order to avoid some problems
that occur in the standard Maxwell theory. The interest for such models has started with the pioneering work of Born and Infeld \cite{BI} whose main motivation was
to modify the standard Maxwell theory in order to eliminate the problem of infinite energy of the electron. Nonlinear electrodynamics is also crucial for the construction of regular black holes \cite{eloy} and for anisotropic black hole solutions \cite{anisoBHnonline}. In addition, owing to their peculiar
thermodynamics properties, nonlinear electrodynamic models have also attracted a lot attention from the physics community \cite{thermoBHnonline}.

In our case, we will see that the appropriate form is given by a power of the Maxwell invariant, namely
$\mathcal{L}_2=(-F_{\mu\nu}F^{\mu\nu})^p$. The presence of the minus sign multiplying the Maxwell invariant ensures the existence of real solutions for any exponent $p$. This model has been intensively studied during the last decade \cite{nonlinMax}. Hence, in the search of charged black hole solutions, we will
consider the following four-dimensional action
\begin{equation}
S=\kappa\int{\sqrt{-g}(R-2\Lambda)}d^4x+\frac{\beta}{4}\int{\sqrt{-g}(-F_{\mu\nu}F^{\mu\nu})^{p}}d^4x-\frac{\alpha}{2}\int{\sqrt{-g}A^2\nabla_\mu A^\mu}d^4x,
\end{equation}
where $\beta$ is assumed to be positive in order to recover the standard Maxwell theory in the limit $p\rightarrow 1$. The variations with respect to the metric tensor and the vector field yield
\begin{equation}
E_{\mu\nu}=\kappa (G_{\mu\nu}+\Lambda g_{\mu\nu})-\frac{\beta}{2}T_{\mu\nu}^{(1)}-\frac{\alpha}{2}T_{\mu\nu}^{(2)}=0
\end{equation}
\begin{equation}
\mathcal{E}^\nu=p\beta  \nabla_{\mu}(\mathcal{F}^{p-1}F^{\mu\nu})-\alpha A^{\nu}\nabla_{\alpha}A^{\alpha}+\frac{\alpha}{2}\nabla^{\nu}(A^2)=0
\end{equation}
where we have defined
\begin{eqnarray}
T_{\mu\nu}^{(1)}&=&p\mathcal{F}^{p-1}F_{\mu\lambda}F_{\nu}^{\lambda}+\frac{1}{4}g_{\mu\nu}\mathcal{F}^{p},\\
T_{\mu\nu}^{(2)}&=&A_{\mu}A_{\nu}\nabla_{\alpha}A^\alpha-\frac{1}{2}A_\mu\nabla_{\nu}(A^2)-\frac{1}{2}A_\nu\nabla_{\mu}(A^2)+\frac{1}{2}g_{\mu\nu}A^\alpha\nabla_{\alpha}A^2,
\end{eqnarray}
and where $\mathcal{F}=-F_{\mu\nu}F^{\mu\nu}$.

Using the Ansatz defined in (\ref{metric}) and (\ref{proca}) with
$N(r)=1$ and setting $a:=A_t$ and $\phi:=A_r$, the equations of
motion become\footnote{ We have set $N(r)=1$ since the solutions
of Einstein gravity with the nonlinear source are only known for
this particular Ansatz.}
\begin{eqnarray}
\label{EqsMotionNonlin}
&&\mathcal{E}_t:=-2a{a^{\prime}}^2(rf^{\prime}+2f)\phi-f\left[\beta p {a^{\prime}}^{2p}{a^{\prime}}^{\prime}r(2^p-p2^{p+1})+
2a{a^{\prime}}^2r\phi^{\prime}-\beta p (2a^{\prime})^{2p+1}\right]=0,\nonumber\\
&& \mathcal{E}_r:=4f+rf^{\prime}\phi^2+ra(af^{\prime}-2a^{\prime}f)=0,\nonumber\\
&& E_{tt}:=2\phi^2r^2f^2(\phi f^{\prime}+2f\phi^{\prime})-[4ar(a^{\prime}r+2a)f+2a^2f^{\prime}r^2]\phi-8\kappa f^2
+f[8\kappa(1-rf^{\prime}-\Lambda r^2)\nonumber\\
&&\qquad \qquad \qquad \qquad \qquad \qquad \qquad -4a^2r^2\phi^{\prime}+\beta {a^{\prime}}^{2p}r^2(2^p-p2^{p+1})]=0,\\
&& E_{rr}:=8 \kappa f^2-2rf(rff^{\prime}-4f)\phi^3-[2ar^2(af^{\prime}-2fa^{\prime})]\phi\nonumber\\
&&\qquad \qquad \qquad \qquad \qquad \qquad \qquad+f\left[8\kappa(rf^{\prime}-1+\Lambda r^2)-\beta {a^{\prime}}^{2p}r^2(2^p-p2^{p+1})\right]=0,\nonumber\\
&& E_{tr}:=rf^{\prime}(a^2-f^2\phi^2)-4\phi^2f^3-2rfaa^{\prime}=0.\nonumber
\end{eqnarray}
The conditions (\ref{cond1}) and (\ref{cond2}) impliy that $a(r)=\frac{Q}{r^2}$ and $\phi(r)=\frac{Q}{f(r)r^2}$,
and, as a direct consequence, the field equations $\mathcal{E}_r$ and $E_{tr}$ are automatically satisfied. The remaining equations take the form
\begin{equation}
\mathcal{E}_t=-\frac{\beta p}{4Q}r^2f\left(\frac{8Q^2}{r^6}\right)^p(6p-5)=0,   \label{maxwellt}
\end{equation}
and
\begin{equation}
E_{tt}=E_{rr}=8\kappa(r f^{\prime}-1+ f+8\Lambda r^2)+\beta r^2 \left(\frac{8Q^2}{r^6}\right)^p(2p-1)=0,   \label{finaleinstein}
\end{equation}
and we see that, in order to satisfy the nonlinear Maxwell equation $\mathcal{E}_t$, we have $p=\frac{5}{6}$. This justifies a posteriori   the minus sign multiplying the Maxwell invariant owing that this latter is negative definite for our Ansatz. Finally,  the
remaining independent equation yields
\begin{equation}
f(r)= \frac{r^2}{l^2}-\frac{M}{r}+1+\beta\frac{\sqrt{2}}{6\kappa}\frac{Q^{\frac{5}{3}}}{r^3},
\end{equation}
where, as usual, we have defined the (A)dS radius
$l^2=-\frac{3}{\Lambda}$ (in the Sitter case, the solution is
still valid and corresponds to an imaginary value of $l$). This
family of solutions is asymptotically (A)dS if $\Lambda\neq0$ and
asymptotically locally flat otherwise. In both case, there is a
curvature singularity at the origin, revealed by
 the scalar curvature which reads
\begin{equation}
R=4\Lambda-\beta\frac{\sqrt{2}}{3\kappa}\frac{Q^{\frac{5}{3}}}{r^5}.
\end{equation}
This singularity is hidden by the horizon(s) located, as usual, at $r_{h}$, such that $f(r_h)=0$.

Note that this solution is easily extended to  $D>4$ dimensions. Indeed, in such a case, the divergenceless and the null condition on the vector field implies that
$$
A_t(r)=\frac{Q}{r^{D-2}},
$$
and the nonlinear electrodynamic term in the Lagrangian is given by $\mathcal{L}_2=(-F_{\mu\nu}F^{\mu\nu})^p$ with
$$
p=\frac{2D-3}{2(D-1)}.
$$

\section{Black hole rotating stealth solutions}

We now show that the static stealth solutions can be generalized to a stealth configuration described by the Kerr metric. We
will follow the same strategy as before  by considering the action  (\ref{Lag}) with $\mathcal{L}_2=0$ and with a vanishing cosmological
constant\footnote{The restriction $\Lambda=0$ is just for simplification and the solution we obtain can easily be generalized to a stealth configuration
on the Kerr (A)dS spacetime. Stealth configurations on rotating black holes have also been found for conformally coupled scalar fields in \cite{Anabalon:2009qt}.}. The usual conditions $\nabla_{\mu}A^{\mu}=0$ and $A_{\mu}A^{\mu}=0$ for the
field equations (\ref{sys1}-\ref{sys2}) with $\mathcal{L}_2=\Lambda=0$, in the case of a stationary and axisymmetric Ansatz, imply that the spacetime metric
solution is given by the Kerr metric.  In Kerr-Schild coordinates this is given by
\begin{align}
ds^{2}  & =-d\bar{t}^{2}+d\bar{r}^{2}+\Sigma d\bar{\theta}^{2}+\left(  \bar
{r}^{2}+a^{2}\right)  \sin^{2}\bar{\theta}d\bar{\phi}^{2}-2a\sin^{2}%
\bar{\theta}d\bar{r}d\bar{\phi}\nonumber\\
& +\frac{2M\bar{r}}{\Sigma}\left(  d\bar{t}+d\bar{r}-a\sin^{2}\bar{\theta
}d\bar{\phi}\right)  ^{2}\ ,
\end{align}
where $\Sigma:=\bar{r}^{2}+a^{2}\cos^{2}\bar{\theta}$, $M$ is the mass and $a=J/M$ is the rotation parameter. We also assume that the non-vanishing components of the
potential $A$ are along the time and radial coordinate, i.e. $A_{\mu}=A_{\bar{t}}\delta_{\mu}^{\bar{t}}+A_{\bar{r}}\delta_{\mu}^{\bar{r}}\ $,
with components being functions of the radial and polar coordinates $\bar{r}$ and $\bar{\theta}$. In this case, the transverse and null conditions imply that
\begin{equation}
A^{\bar{r}}(\bar{r},\bar{\theta})=\frac{F\left(  \bar{\theta}\right)  }{\bar{r}^{2}+a^{2}\cos
^{2}\bar{\theta}}\ ,
\end{equation}
and this restriction is consistent with two different expressions of the scalar potential that are given by
\begin{equation}
A^{\bar{t}}(\bar{r},\bar{\theta})=-\frac{F\left(  \bar{\theta}\right)  }{\bar{r}^{2}+a^{2}\cos
^{2}\bar{\theta}}\ ,\label{At1KS}
\end{equation}
and
\begin{equation}
A^{\bar{t}}(\bar{r},\bar{\theta})=\frac{F\left(  \bar{\theta}\right)  \left(  \bar{r}^{2}+2M\bar
{r}+a^{2}\cos^{2}\bar{\theta}\right)  }{\left(  \bar{r}^{2}+a^{2}\cos^{2}%
\bar{\theta}\right)  \left(  \bar{r}^{2}-2M\bar{r}+a^{2}\cos^{2}\bar{\theta
}\right)  }\ .\label{At2KS}
\end{equation}
In these equations $F\left(\bar{\theta}\right)$ is an arbitrary function of the polar coordinate. In order to better clarify these two
classes of rotating stealth configurations, we write them in Boyer-Lindquist coordinates $\left(t,r,\theta,\phi\right)$ defined by
\begin{align}
t  & =\bar{t}-\int\frac{2M\bar{r}}{\bar{r}^{2}-2M\bar{r}+a^{2}}d\bar{r}\ ,\\
\phi & =\bar{\phi}-\int\frac{a}{\bar{r}^{2}-2M\bar{r}+a^{2}}d\bar{r}\ .
\end{align}
In this coordinate system, the metric reads
\begin{eqnarray*}
ds^2&=&-dt^2+\rho(r,\theta)^2\left(\frac{dr^2}{\Delta(r)}+d\theta^2\right)+\left(r^2+a^2\right)\sin(\theta)^2d\phi^2+
\frac{2Mr}{\rho(r,\theta)^2}\left(a\sin(\theta)^2d\phi-dt\right)^2
\end{eqnarray*}
where $\rho(r,\theta)^2=r^2+a^2\cos(\theta)^2$ and $\Delta(r)=r^2-2Mr+a^2$. It is worth to note that through
the coordinate change, the potential $A_{\mu}$ acquires an  additional component, and the first class of rotating stealth configuration (\ref{At1KS}) becomes in the,
Boyer-Lindquist coordinates,
\begin{align*}
A^{t}(r,\theta)  & =-\frac{F\left(  \theta\right)  \left(  r^{2}+a^{2}\right)  }{\left(
r^{2}+a^{2}\cos^{2}\theta\right)  \left(  r^{2}-2Mr+a^{2}\right)  },\\
A^{r}(r,\theta)  & =\frac{F\left(  \theta\right)  }{r^{2}+a^{2}\cos^{2}\theta},\\
A^{\phi}(r,\theta)  & =-\frac{aF\left(  \theta\right)  }{\left(  r^{2}-2Mr+a^{2}\right)
\left(  r^{2}+a^{2}\cos^{2}\theta\right)  },
\end{align*}
while the second class (\ref{At2KS}) is expressed as
\begin{align*}
A^{t}(r,\theta)&=\frac{\left(  r^{2}+a^{2}\right)  \left(  r^{2}+a^{2}\cos^{2}%
\theta\right)  -2rm\left(  r^{2}-a^{2}+2a^{2}\cos^{2}\theta\right)  }{\left(
r^{2}-2Mr+a^{2}\right)  \left(  r^{2}+a^{2}\cos^{2}\theta\right)  \left(
r^{2}-2Mr+a^{2}\cos^{2}\theta\right)  }F\left(  \theta\right),   \\
A^{r}(r,\theta)  & =\frac{F\left(  \theta\right)  }{r^{2}+a^{2}\cos^{2}\theta},\\
A^{\phi}(r,\theta)  & =-\frac{aF\left(  \theta\right)  }{\left(  r^{2}-2Mr+a^{2}\right)
\left(  r^{2}+a^{2}\cos^{2}\theta\right)  }.
\end{align*}
A way for distinguishing these two classes of solutions is to remark that the t-component of the last stealth configuration
diverges at the ergosphere. Hence, it may be reasonable to consider the first configuration as the physically sound one.

\section{Conclusions and further research}
In this work, we have found black hole solutions of the generalized Proca field theory exhibiting nonminimal couplings of the vector field with the
curvature along with self-derivatives interactions. We focused on the simplest Lagrangian contained in (\ref{generalizedProfaField_curved})
and given by the action (\ref{Lag}). For this model, we have shown, using a similar argument to the no-hair theorem of Bekenstein, that the transverse
condition $\nabla_\mu A^\mu =0$ must be associated with the null condition $A_\mu A^\mu =0$. As shown in this work, these two conditions are very
helpful in our search of black hole solutions. Firstly, in the absence of the $\mathcal{L}_2$ term in the action (\ref{Lag}), these two conditions necessarily
imply that the solution must be a stealth configuration. In this case, we have obtained two non-trivial black hole stealth configurations defined on Schwarzschild
and on the Kerr spacetimes. The scalar potential in the static case does not exhibit a Coulombian behavior, and instead behaves as $Q/r^2$. We have taken inspiration from the nonlinear electrodynamic models that admit non-Coulombian fields to chose appropriately the Lagrangian $\mathcal{L}_2$ in order to construct black hole solutions that differ from (A)dS black holes.

Some further directions can be followed to extend our analysis. For example, it will be interesting to explore more general truncations
of the full Lagrangian (\ref{generalizedProfaField_curved})  and the corresponding transverse and null conditions that yield to stealth configurations. Also the
analysis of the thermodynamical properties of our new solutions are worth invesitgating. Indeed, for the stealth  black holes, the solutions are
characterized by an integration constant $Q$ in addition to the standard mass parameter $M$. It will be nice to clarify the physical interpretation of this constant,
which, in the nonlinear case, is proportional to the electric charge. Finally, as a very nontrivial task, it will be interesting to find the rotating version of the static solution in the nonlinear electrodynamic case.

\section{Acknowledgments}
A.C would like to express his gratitude to the Theoretical Physics Department of the University of Geneva for its kind hospitality during the initial stage of this work.
This work is partially supported by grants 1130423, 1141073 and 3150157, from FONDECYT. This project is also partially funded by Proyectos CONICYT- Research Council UK - RCUK
-DPI20140053. A.C. and J. O. would like to thank the International Center for Theoretical Physics (ICTP), Trieste, Italy, where part of this work was carried out.
\section{Appendix: Higher dimensional rotating stealths}
The rotating stealth solution can be further extended to higher dimensions. In fact, the extension of the Kerr metric to dimensions greater than four was found by
Myers and Perry in \cite{Myers:1986un}. The spacetime describes a black hole characterized by
$\left[  \frac{D-2}{2}\right]  $ rotation parameters $a_{i}$, and a mass
parameter $M$. The metric can be written in a Kerr-Schild form and provided
$D=N+1$ we have, for odd $N$ that the metric reads
\begin{align}
ds^{2}  & =-dt^{2}+dr^{2}+r^{2}d\alpha^{2}+\left(  r^{2}+a_{i}^{2}\right)
\left(  d\mu_{i}^{2}+\mu_{i}^{2}d\phi_{i}^{2}\right)  +2a_{i}\mu_{i}^{2}%
d\phi_{i}dr\nonumber\\
& +\frac{Mr^{2}}{\Pi F}\left(  dt+dr+a_{i}\mu_{i}^{2}d\phi_{i}\right)  ^{2}\ ,\nonumber
\end{align}
where $\alpha^{2}=1-\mu_{i}^{2}$ and where the sum is understood for repeated index $i$ with
$i=1,...,\frac{N-1}{2}$. The functions $F$ and $\Pi$ are functions that depend on $r$ and $\mu_{i}$. For
even $N$ the metric solution reads
\begin{align}
ds^{2}  & =-dt^{2}+dr^{2}+\left(  r^{2}+a_{i}^{2}\right)  \left(  d\mu_{i}%
^{2}+\mu_{i}^{2}\right)  +2a_{i}\mu_{i}^{2}d\phi_{i}dr+K\left(  r,\mu_{i}\right)  \left(  dt+dr+a_{i}\mu_{i}^{2}d\phi_{i}\right)
^{2}\ ,\nonumber
\end{align}
where now the coordinates $\mu_{i}$ are restricted such that $\mu_{i}^{2}=1$. In both cases (even and odd), the expressions for $\Pi$ and $F$ read
\begin{align}
F  & =1-\frac{a_{i}^{2}\mu_{i}^{2}}{r^{2}+a_{i}^{2}}\ ,\nonumber\\
\Pi & ={\displaystyle\prod\limits_{i=1}^{\left(  N-1\right)  /2}}
\left(  r^{2}+a_{i}^{2}\right)  \ ,\nonumber
\end{align}
while the function $K$ is given by
\[
K\left(  r,\mu_{i}\right)  =\left\{
\begin{array}
[c]{cc}%
\frac{Mr}{\Pi F} & \text{for odd }N,\\
\frac{Mr^{2}}{\Pi F} & \text{for even }N.
\end{array}
\right.
\]
We are concerned with the problem of finding null and divergenless
vectors on these spacetimes. Assuming as in the four-dimensional case, the Ansatz  $A^{\mu}=A^{t}\left(
r,\mu_{i}\right)  \delta_{t}^{\mu}+A^{r}\left(  r,\mu_{i}\right)  \delta
_{r}^{\mu}$, the following configuration provides a stealth generalized Proca
field on the Myers-Perry background
\begin{equation}
A^{r}=\frac{C\left(  \mu_{i}\right)  }{\sqrt{-g}}\ ,
\end{equation}
where $C$ is an arbitrary function of the angles $\mu_{i}$ ; this
expression is compatible with two possible solutions  of the $t$-component of
$A^{\mu}$ given by
\begin{equation}
A^{t}=-A^{r}\ ,
\end{equation}
or
\begin{equation}
A^{t}=\left(  \frac{1+K}{1-K}\right)  A^{r}\ .\label{sbd}
\end{equation}
As before, the metric can be transformed into Boyer-Lindquist coordinates where the
vector field will also acquire components along the $\phi_{i}^{2}$ direction,
giving rise to  magnetic components.

\end{document}